\documentclass[aip,apl,reprint,groupedaddress,preprintnumber,twocolumn,showpacs]{revtex4-1}
\usepackage{amsmath}
\usepackage{amssymb}
\usepackage{graphicx}
\usepackage{amssymb}
\usepackage{graphics}
\usepackage{graphicx}
\usepackage{epsfig}
\usepackage{color}

\begin{document}


\title{Three-dimensional Fermi surface and small effective masses in Mo$_8$Ga$_{41}$}
\author{Zhixiang Hu,$^{1,2,\ddag}$ D. Graf,$^{3}$ Yu Liu,$^{1}$ and C. Petrovic$^{1,2,\ddag}$}
\affiliation{$^{1}$Condensed Matter Physics and Materials Science Department, Brookhaven National Laboratory, Upton, New York 11973, USA\\
$^{2}$Department of Physics and Astronomy, Stony Brook University, Stony Brook, New York 11794-3800, USA\\
$^{3}$National High Magnetic Field Laboratory, Florida State University, Tallahassee, Florida 32306-4005, USA}

\date{\today}

\begin{abstract}
We report Fermi surface characteristics of Mo$_{8}$Ga$_{41}$, a two-gap superconductor with critical temperature $T_{c}$$\sim$10 K, obtained from quantum oscillation measurements. Four major frequencies have been observed with relatively small quastiparticle masses. Angular dependence of major frequencies indicates three-dimensional Fermi surface sheets. This argues for a relatively isotropic superconducting state and, given its relatively high $T_c$, shows that a search for materials in this class could be of interest for superconducting wire applications.

\end{abstract}

\maketitle


Magnesium diboride with its superconducting critical temperature $T_c$ similar to doped La$_{2}$CuO$_{4}$, rather simple crystallography and phonon-mediated superconductivity stimulated considerable interest.\cite{Nagamatsu,Budko} Indeed, recent years have reaffirmed that room-temperature superconductors should be sought among non-magnetic materials with high phonon frequencies\cite{Drozdov1,Drozdov2,Errea} rather than in magnetic copper oxide ceramics and spin-fluctuation pairing mechanism.\cite{Scalapino}

Similar to MgB$_{2}$, Mo$_{8}$Ga$_{41}$ is a binary intermetallic superconductor. Its superconducting $T_{c}$ = 9.7 K and upper critical field $\mu$$_{0}$$H_{c2}$ = 8.36 T.\cite{BezingeA} It has been pointed out that, like MgB$_{2}$, Mo$_{8}$Ga$_{41}$ is also on the verge of a structural instability,\cite{Slusky,Xie} exhibits strong-coupling two-gap BCS superconductivity and good flux pinning potential.\cite{Verchenko,Verchenko2,Neha,Sirohi,Marcin,Larbalastier} In light of the relatively high critical temperature and upper critical field, further studies of this material are highly desirable.

\begin{figure}
\centerline{\includegraphics[scale=0.4]{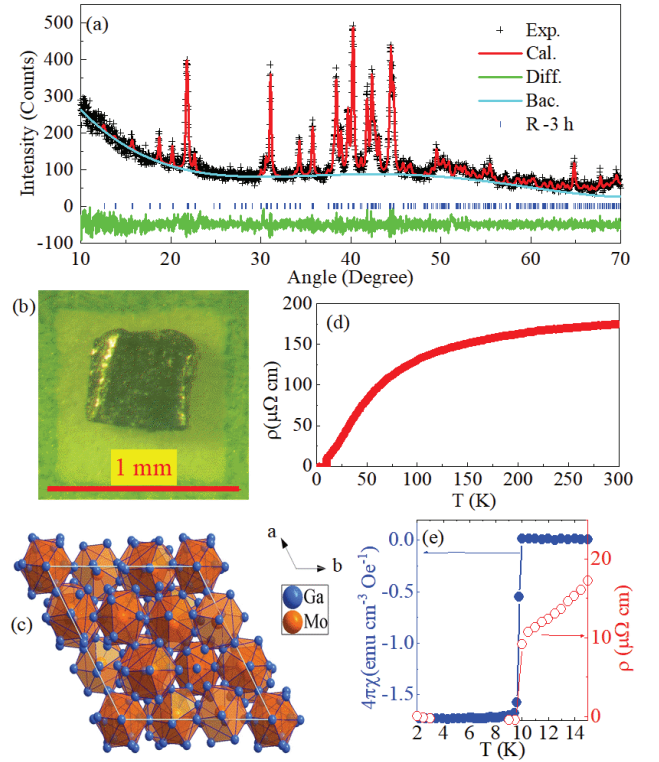}}
\caption{(Color online). (a) Powder XRD pattern and refinement results. The data were shown by (+), fitting, difference and background curves are given by the red, green and blue solid lines, respectively. Single crystals (b) and crystal structure (c) of Mo$_{8}$Ga$_{41}$. (d) Temperature dependence of the in-plane resistivity. (e) Superconducting $T_c$ measured by in-plane magnetic susceptibility in a magnetic field of 10 Oe and zero-field resistivity indicating $T_c$ = 9.8(1) K.}
\label{res}
\end{figure}

Fermi surface characteristics give insight into conducting states that are subject to Cooper pairing and are rather important for predictive materials design.\cite{BergemannC,SebastianS,SebastianS2,ProustC} In this work we have performed a comprehensive quantum oscillations study of Mo$_{8}$Ga$_{41}$. Multiple Fermi surface sheets with three-dimensional character have been found with rather small quasiparticle masses. Three-dimensional Fermi surfaces and two-band superconductivity argue in favor of further search for materials in this class for superconducting wire applications.\cite{Buzea,JinS}

Single crystals of Mo$_8$Ga$_{41}$ were grown by high temperature self-flux method.\cite{Fisk} Mo and Ga were mixed in a ratio of 8:500 in an alumina crucible and sealed in evacuated quartz tube. The ampoule was heated up to 850$^\circ$ in two hours, held at 850 $^\circ$C for 10 h, and then slowly cooled to 170 $^\circ$C for 55 hours, where crystals were decanted. Residual gallium was removed by crystal etching in diluted hydrochloric acid. X-ray diffraction (XRD) was performed on crushed crystals at room temperature by using Cu-K$_\alpha$ ($\lambda$ = 0.15418 nm) radiation in a Rigaku Miniflex powder diffractometer. Chemical composition was verified by the energy-dispersive x-ray spectroscopy (EDX) in JEOL LSM-6500 scanning electron microscope. Magnetic susceptibility was measured in the zero-field cooling mode in a Quantum Design MPMS-5XL. The de Haas van Alphen (dHvA) oscillation experiments were performed at NHMFL Tallahassee. The crystals were mounted onto miniature Seiko piezoresistive cantilevers which were installed on a rotating platform. The field direction can be changed continuously between parallel ($\theta$ = 0$^{\circ}$) and perpendicular ($\theta$ = 90$^{\circ}$) to the $c$-axis of the crystal. The average inverse magnetic field was determined from (H$^{-1}_{max}$  +H$^{-1}_{min}$)/2.

Powder X-ray diffraction (XRD) pattern [Fig. 1(a)] confirms that single crystals crystallize in a \textit{R -3h} space group. The refined lattice parameters are $a$ = $b$ = 1.4031(2)nm and $c$ = 1.5042(2) nm in agreement with reported values.\cite{Yvon} Fig. 1(b) shows typical crystal size, about two times larger than previously reported.\cite{Verchenko} EDX shows a stoichiometric ratio of 0.167:0.833, which is close to the Mo:Ga ratio of 8:41. Crystal structure of Mo$_{8}$Ga$_{41}$ consists of stacked MoGa$_{10}$ Mo-Ga polyhedral cages with Mo atoms at the center, with two different atomic positions for Mo and nine for Ga [Fig. 1(c)].\cite{Verchenko,Yvon} Resistivity shows the absence of phase transition above superconducting $T_c$, residual resistivity ratio RRR = $\rho$(300 K)/$\rho$(10 K) = 17 and $\rho$(10 K) = 10 $\mu$$\Omega$ cm [Fig. 1(d)]. Superconducting T$_c$'s inferred from the diamagnetic signal of our crystals [Fig. 1(e)] is consistent with polycrystaline Mo$_8$Ga$_{41}$.\cite{BezingeA}

\begin{figure}
\centerline{\includegraphics[scale=0.35]{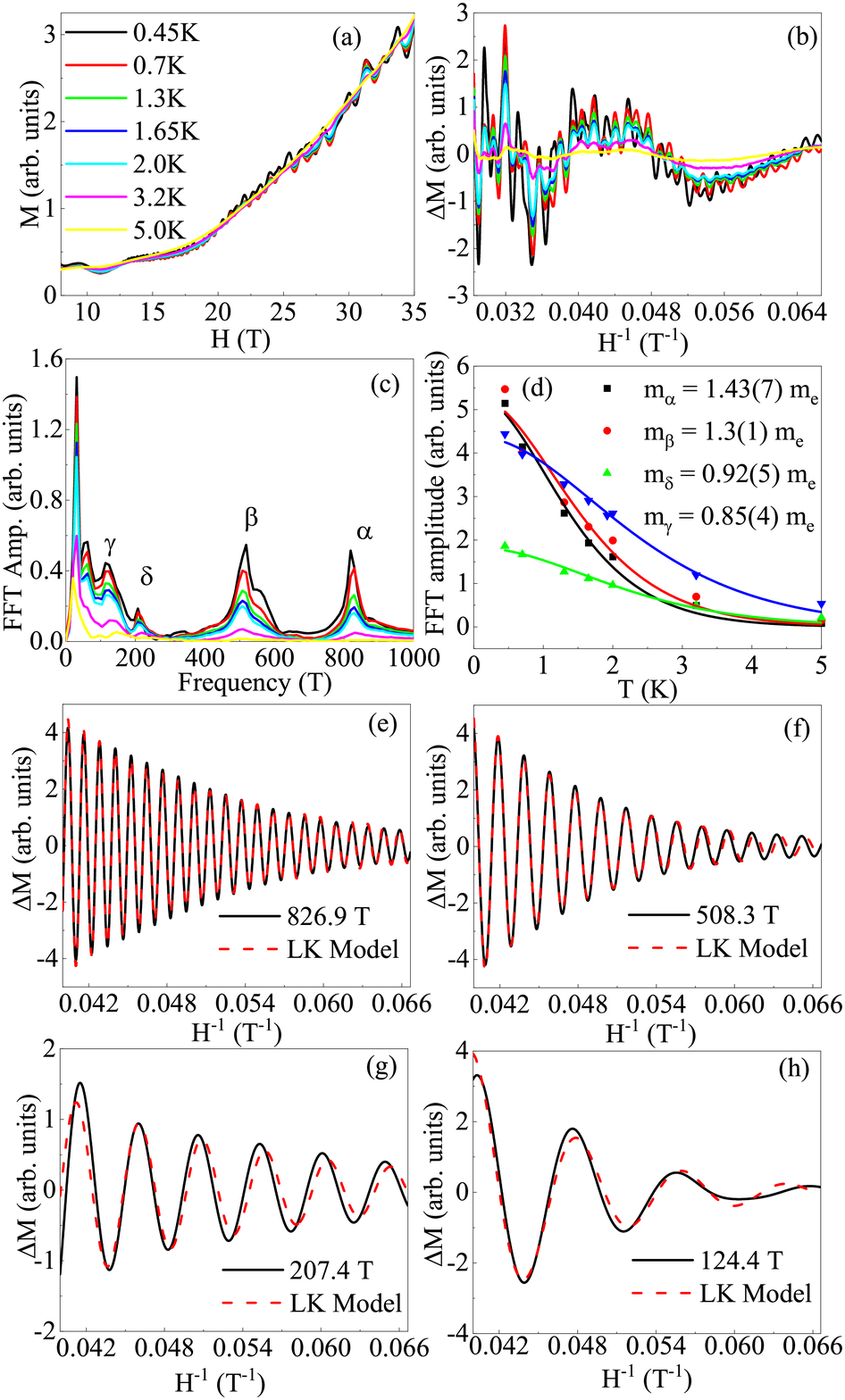}}
\caption{(Color online). (a) Temperature dependent cantilever response of Mo$_8$Ga$_{41}$ versus magnetic field, with the field applied along $c$ axis. (b) Oscillatory component obtained by smooth background subtraction. (c) FFT spectrum of oscillations vs. frequency with observed frequencies $\alpha$, $\beta$, $\gamma$ and $\delta$ at different temperatures. Their effective masses are evaluated by using Lifshitz-Kosevich (LK) formula (d). (e-h) Magnetic field-dependent amplitudes of characteristic frequencies.}
\label{res}
\end{figure}

\begin{figure}
\centerline{\includegraphics[scale=0.35]{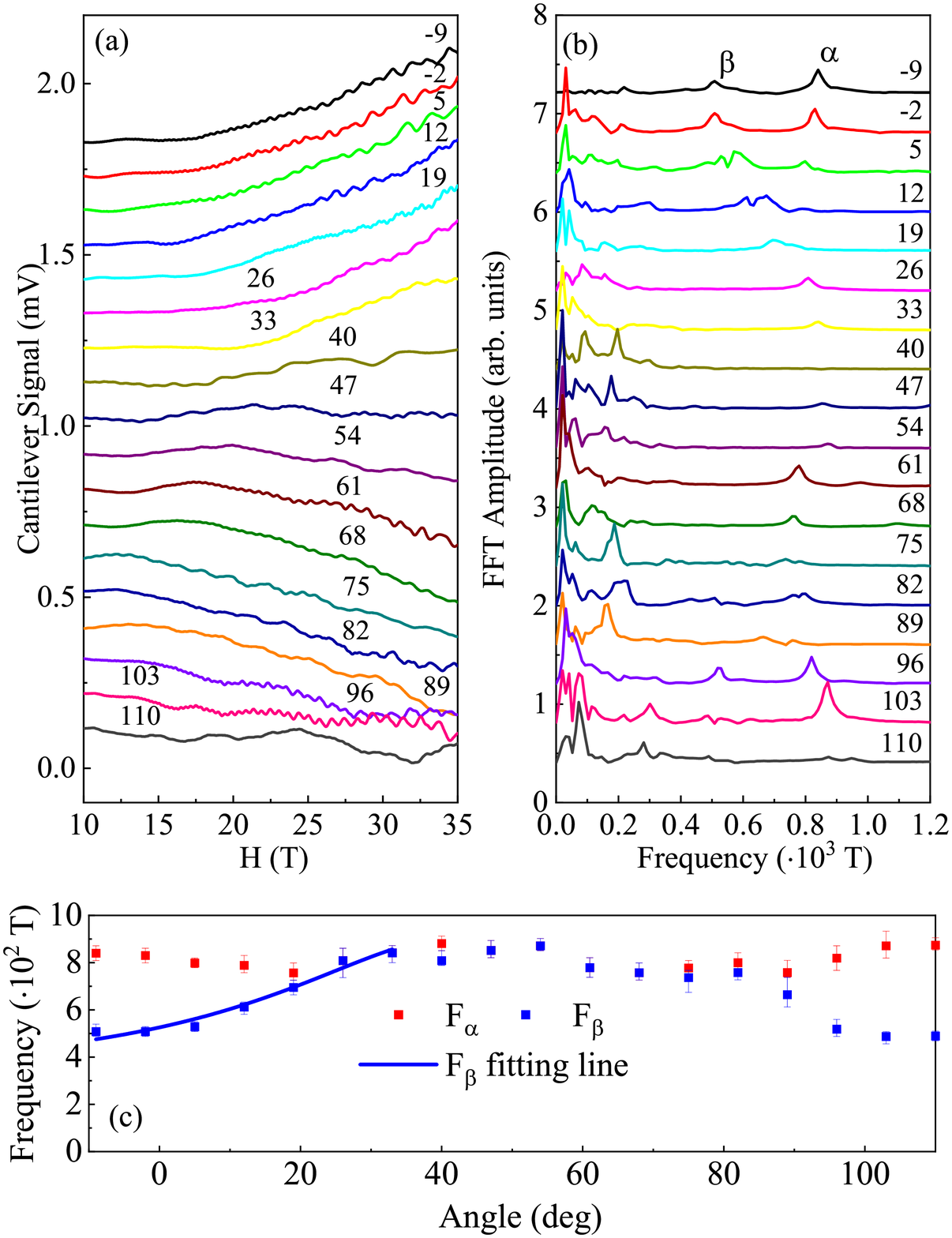}}
\caption{(Color online). (a) Angle-dependent dHvA oscillations versus magnetic field with offset for clarity. The field was applied along $c$ axis and the sample was rotated from -9$^\circ$ to 110$^\circ$; (b) FFT of response $\Delta V = V - \langle V \rangle$ versus Frequency at different angles (c) Traces of frequencies $\alpha$, and $\beta$ with the angle fitted with the ellipsoid model}.
\label{magnetism}
\end{figure}

Field-dependent cantilever signal [Fig. 2(a)] in magnetic field applied along the $c$-axis shows clear de Haas van Alphen (dHvA) oscillations above 10 T. Cantilever signal and dHvA amplitude increase in magnetic field at all temperatures. Oscillating part of the signal and its Fast Fourier Transform (FFT) [Fig. 2 (b,c)] show sharp peaks. The peaks correspond to the orthogonal cross-sectional area of the Fermi surface (FS) $A_F$ as described by the Onsager relation $F$ = ($\Phi_0$/2$\pi^2$)$A_F$, where $\Phi_0$ is the flux quantum. Therefore we estimate $A_{F\alpha}$ = 7.88(1) nm$^{-2}$, $A_{F\beta}$ = 4.85(1) nm$^{-2}$, $A_{F\gamma}$ = 1.18(1) nm$^{-2}$, $A_{F\delta}$ = 1.98(1) nm$^{-2}$ for frequencies $F_\alpha$ = 826.9 T, $F_\beta$ = 508.3 T, $F_\gamma$ = 124.4 T and $F_\delta$ = 204.7 T, respectively. Effective cyclotron mass of carriers at observed Fermi surface sheets is obtained from the temperature dependence of FFT amplitude from the Lifshitz-Kosevich (LK) formula:\cite{Lifshitz,Shoenberg}
\[FFT\ amp. \propto \dfrac{\alpha m^*T/H}{\sinh(\alpha m^*T/H)}, \]
where $\alpha = 2\pi^2K_B/e\hbar \approx 14.69$ T/K and $m^* = m/m_e$ is the effective cyclotron mass. The effective masses are estimated [Fig. 2(d)] to be $1.43(7)m_e$, $1.3(1)m_e$, $0.85(4)m_e$, and $0.92(5)m_e$ for Fermi surface parts that correspond to F$_{\alpha}$, F$_{\beta}$, F$_{\gamma}$ and F$_{\delta}$, respectively.

We obtain Dingle temperature by using Lifshitz-Kosevich formula to fit the field-dependent amplitudes at 0.45 K for oscillation frequencies.\cite{Shoenberg,XiaW,XiaW2} This gives [Fig. 1(e-h)]. $T_{D\alpha}$ = 3.00(1) K, $T_{D\beta}$ = 5.02(3) K, $T_{D\gamma}$ = 8.38(8) K, $T_{D\delta}$ = 3.2(1) K.

Dingle temperature $T_D$ is associated the with mean free path and the scattering rate $\tau$ via $T_{D} = \frac{\hbar}{2\pi{k_B}\tau_Q}$ that describes scattering of carriers due to particle interactions and defects inside the material. Differences in $T_D$ among characteristic frequencies imply different quasiparticle lifetimes of Fermi surface sheets: $\tau_{\alpha}$ = 4.0(1)$\cdot$10$^{-13}$ s, $\tau_{\beta}$ = 2.4(1)$\cdot$10$^{-13}$ s, $\tau_{\gamma}$ = 1.4(1)$\cdot$10$^{-13}$ s and $\tau_{\delta}$ = 3.8(1)$\cdot$10$^{-13}$ s. We can also approximate Fermi wave vector $k_{F,\alpha}$ = 0.158(1) ${\AA}$$^{-1}$, $k_{F,\beta}$ = 0.124(1) ${\AA}$$^{-1}$, $k_{F,\gamma}$ = 0.061(1) ${\AA}$$^{-1}$ and $k_{F,\delta}$ = 0.079(1) ${\AA}$$^{-1}$ assuming circular cross-section $A_F$ = $\pi$$k_F^2$. Mobility estimate using $\mu$ = $e\tau/m_c$ gives 498.0 cm$^2$V$^{-1}$s$^{-1}$, 327 cm$^2$V$^{-1}$s$^{-1}$, 300.0 cm$^2$V$^{-1}$s$^{-1}$ and 726.0 cm$^2$V$^{-1}$s$^{-1}$ for $F_\alpha$, $F_\beta$, $F_\gamma$ and $F_\delta$, respectively.

Angular-dependent dHvA oscillations were also measured with field initially set in the $ab$-plane. The crystal was rotated from -9$^\circ$ to 110$^\circ$.  The dHvA oscillations appear [Fig. 3(a)] at all angles above 10 T. Fig.3(b) shows FFT of dHvA response. Traces of two frequencies $\alpha$, and $\beta$ are clearly resolved and are plotted in Fig.3(c). Frequency $\alpha$ shows oscillatory behavior with maxima at 45$^\circ$ and 110$^\circ$. Frequency $\beta$ increases through small angles ($<30^\circ$), reaches maximum also around 45$^\circ$ and then decreases. Angular-dependence of $F_\beta$ can be described by the ellipsoidal fit\cite{LawsonB} $F(\theta)$ = $F_0$[cos$^2$($\theta$) + ($k^{a}_{F}$/$k^{c}_{F}$)$^2$sin$^2$($\theta$)]$^{-1/2}$ [Fig. 3(c)]. The ratio of ($k^{a}_{F}$/$k^{c}_{F}$) = 0.42(3) describes eccentricity of the Fermi surface sheet associated with $F_\beta$.

The three-dimensional (3D) crystal structure of Mo$_{8}$Ga$_{41}$ should result in rather dispersive bands in momentum space. Interestingly, band structure of Mo$_{8}$Ga$_{41}$ features narrow band dispersion whereas the density of states at the Fermi level is dominated by Mo 4$d$ and Ga4$p$ orbital hybridization.\cite{Neha} The 3D character of dHvA oscillations is consistent with first principle calculations\cite{Sirohi} where four different bands cross the Fermi level. Based on the effective masses obtained in our experiment, it is reasonable to $\alpha$ and $\beta$ frequencies with electron pocket near the Brillouen zone corner and large anisotropic part of the Fermi surface, respectively whereas other frequencies probably arise due to band associated with two hole pockets near the  $\Gamma$ point that exhibit weak coupling with phonons and where superconducting gap is smaller.\cite{Sirohi}  Therefore, bands corresponding to $\alpha$ and $\beta$ Fermi surface sheets that exhibit larger superconducting gap feature three-dimensional character.

In conclusion, we present experimental study of Fermi surface in Mo$_8$Ga$_{41}$ by quantum oscillations. Our results indicate three-dimensional character of bands that exhibit strong coupling with phonons and contribute to larger gap below superconducting $T_{c}$. Given relatively high superconducting critical temperature, further studies of superconducting state and search for similar materials are highly desirable.

This work was supported by the U.S. DOE-BES, Division of Materials Science and Engineering, under Contract No. DE-SC0012704 (BNL). Work at the National High Magnetic Field Laboratory is supported by the NSF Cooperative Agreement No. DMR-1157490, and by the state of Florida.

The data that support the findings of this study are available from the corresponding author upon reasonable request.

$\ddag$ petrovic@bnl.gov and zhixiang@bnl.gov

\end{document}